\documentclass[aps,prb,twocolumn,showpacs,superscriptaddress,10pt,showpacs,amsmath,amssymb,floatfix]{revtex4-1}
\usepackage{graphicx}
\usepackage{amsmath}
\usepackage{amssymb}
\usepackage[utf8]{inputenc}
\usepackage{comment}
\usepackage{dcolumn}
\usepackage{bm}
\usepackage{hyperref}
\usepackage{verbatim}
\usepackage{bbold}

\newcommand{\Tb}{ {\bm T } }
\newcommand{\ch}{\hat{c}}
\newcommand{\chd}{\hat{c}^\dagger}
\newcommand{\Hh}{\hat{H}}

\newcommand{\Gb}{ {\bm G } }
\newcommand{\vb}{ {\bm v } }
\newcommand{\Sigmab}{ {\bm \Sigma } }
\newcommand{\Gammab}{ {\bm \Gamma } }
\newcommand{\nh}{\hat{n}}
\newcommand{\Tr}{\text{Tr}}
\newcommand{\Eq}[1]{Eq.~(\ref{#1})}

\renewcommand\Im{\operatorname{Im}}
\newcommand{\eff}{\text{eff}}
\newcommand{\KS}{\text{KS}}
\newcommand{\Hxc}{\text{Hxc}}
\newcommand{\xc}{\text{xc}}
\newcommand{\MB}{\text{MB}}

\graphicspath{{pics/}}

\begin{document}

  \title{Disorder and interactions in systems out of equilibrium: \\The exact independent-particle picture from density functional theory}

  \author{Daniel Karlsson}
  \affiliation{Department of Physics,
    Nanoscience Center P.O.Box 35 FI-40014 University of Jyv\"{a}skyl\"{a}, Finland}
  \email[Corresponding author:] {daniel.l.e.karlsson@jyu.fi}

  \author{Miroslav Hopjan}
  \affiliation{Department of Physics, Division of Mathematical Physics, Lund University, 22100  Lund, Sweden}
  \affiliation{European Theoretical Spectroscopy Facility, ETSF}

  \author{Claudio Verdozzi}
  \affiliation{Department of Physics, Division of Mathematical Physics, Lund University, 22100  Lund, Sweden}
  \affiliation{European Theoretical Spectroscopy Facility, ETSF}

  \begin{abstract}
    \noindent Density functional theory (DFT) exploits an independent-particle-system construction to replicate the densities and current of an interacting system. This construction is used here to access the exact effective potential and bias of non-equilibrium systems with disorder and interactions. Our results show that interactions smoothen the effective disorder landscape, but do not necessarily increase the current, due to the competition of disorder screening and effective bias. This puts forward DFT as a diagnostic tool to understand disorder screening in a wide class of interacting disordered systems.
  \end{abstract}


  \maketitle
  \section{Introduction}\label{Introduction}
  How disorder and electron correlations shape material properties is a major question of current condensed matter research \cite{Imada1998}. The interest in this problem is many decades old \cite{Wigner1934,Boer1937,Mott1937,Mott1949,Korringa1958,Anderson1958,Edwards1972}, and significant progress has been made in important directions, e.g. in describing  the correlation-induced Mott-Hubbard \cite{Gebhard2000} and the disorder-induced Anderson \cite{Abrahams1979} metal-insulator transition. Yet, a complete general understanding of the joint effect of interactions and disorder remains elusive to this day.

  Advances in ultracold-atoms experiments \cite{Lewenstein2007,Bloch2012} have boosted interest in scenarios where disorder and interactions are simultaneously important and new implications emerge from their interplay.
  An example of recently observed phenomena \cite{Choi2016} is many-body localization (MBL) \cite{Gornyi2005,Basko2006a}, a new experimental and theoretical paradigm where several notions of many-body physics blend coherently \cite{Nandkishore2015}. In fact, MBL is part of a broad palette of situations. For example, disorder or interactions alone can produce insulating behavior but, between these limits, how they simultaneously affect conductance is not fully settled \cite{Kramer1993a,Kravchenko1994,Kravchenko1995,Kravchenko1996,Vojta1998,Scalettar2007,Lahoud2014}.  In equilibrium, interactions can increase or decrease conductivity in a disordered system \cite{Denteneer1999,Vojta1998,Abrahams2001}. Out of equilibrium, results for quantum rings \cite{Farhoodfar2011} and quantum transport setups~\cite{Karlsson2014b} suggest that at fixed disorder strength the current depends non-monotonically on interactions.

  To facilitate the description of disordered and interacting systems, it would be useful to have a simple picture. A recent example in this direction was to look at a reduced quantity, the one-body density matrix, to establish a link between MBL,  Anderson localization and Fermi-liquid-type features in closed systems \cite{Bera2015,Bera2017}.
  Another possible reduced description would be in terms of an {\it independent-particle Hamiltonian}.  In a traditional mean-field-intuitive description of disorder vs interactions \cite{Scalettar2007}, the low-energy pockets of the rugged potential landscape attract high particle density, but this is opposed by inter-particle repulsion, resulting in a flatter effective potential landscape, i.e. disorder is screened by interactions. It is not unambiguous how to define such potential, and different conclusions are reached in the literature \cite{Tanaskovic2003, Henseler2008, Henseler2008a, Song2008, Tanaskovic2003, Pezzoli2010, Farhoodfar2011}.
  The question is even more delicate for open systems, where typical localization signatures are unavailable \cite{Luschen2017}. As such, a simple, rigorous picture valid also in the presence of reservoirs would be of utmost importance.

  \begin{figure} [t]
    \centering
    \includegraphics[width=0.47\textwidth]{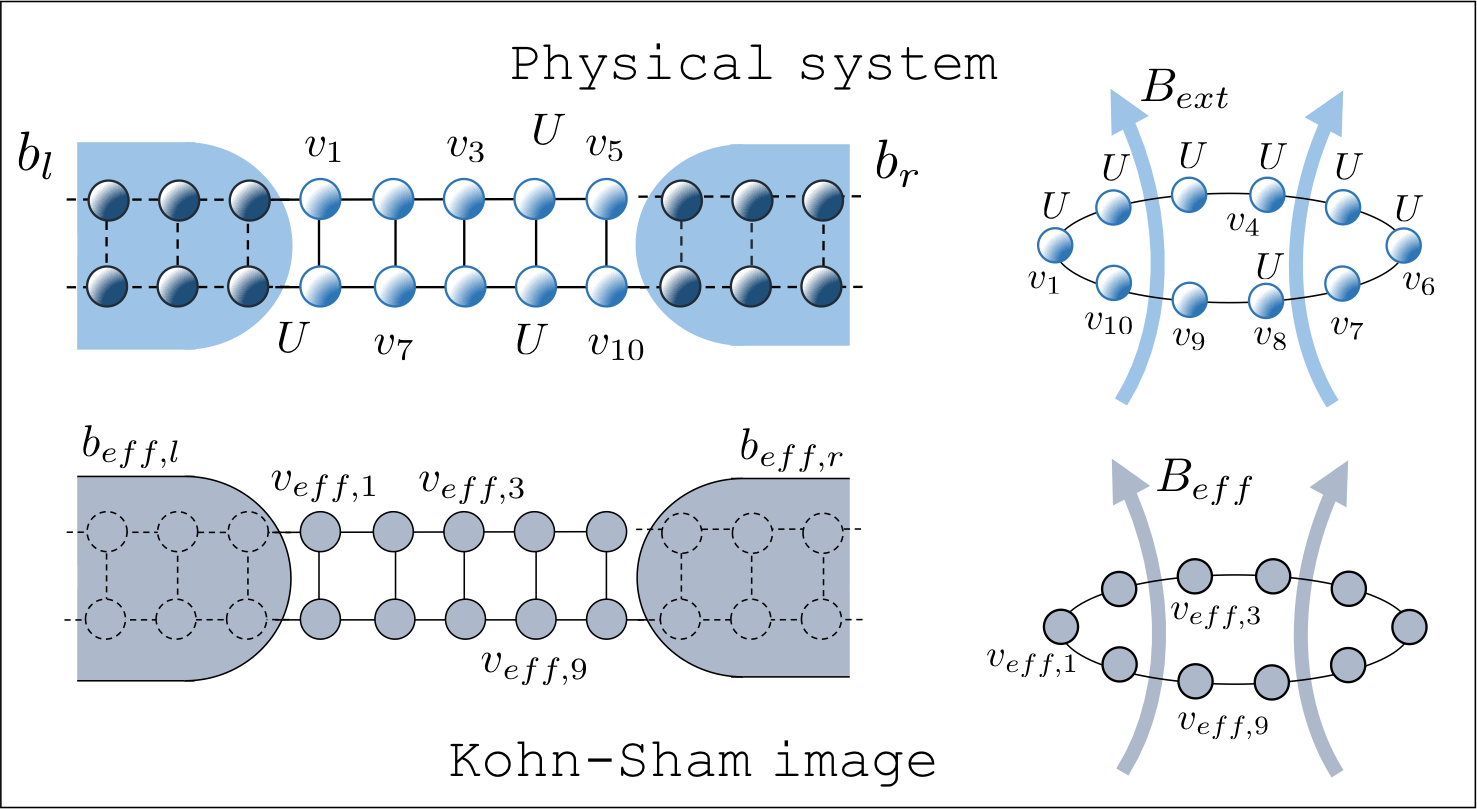}
    \caption{Many-body and corresponding Kohn-Sham systems for rings and 2D quantum transport setups. The interaction $U$, the one-body potentials $\{v_i \}$ and KS potentials $\{v_{\eff,i}\}$ are shown at representative sites.
      \label{system} }
  \end{figure}

  Motivated by these arguments, we introduce here a picture of disorder and interactions based on the Kohn-Sham (KS) independent-particle scheme \cite{Kohn1965} of density functional theory (DFT) \cite{Hohenberg1964,Runge1984}.
  In DFT, the exact density of the interacting system can be obtained from a KS system subjected to an effective potential $v_{\eff}$ (Fig.~\ref{system}). For the density, $v_{\eff}$ is the best effective potential in an independent-particle picture. We propose  that  $v_{\eff}$ {\it can be identified as the independent-particle effective energy landscape in a disordered and interacting system}, which unambiguously defines disorder screening. To assess disorder screening for conductance and currents, we consider out-of-equilibrium systems. In  extending DFT to non-equilibrium, we also have to include the notion of an {\it effective bias} \cite{Schmitteckert2013,Stefanucci2015, Karlsson2016a}.

  Our main findings are:
  i) interactions smoothen the effective landscape seen by the electrons (we interpret this as disorder screening);
  ii) a non-monotonic dependence of the current on the interaction strength cannot be explained by disorder screening alone; an ``effective bias"  (corresponding
  to a screening of the applied bias due to electron correlations) has to be taken into account; iii) The picture from i) and ii) applies to both isolated and open systems and to different dimensionality;
  iv) More in general, our works paves the way to a rigorous understanding of the notion of effective disorder \cite{Patel2017} in a variety of situations,
  including open systems in-and out-of equilibrium, a topic which is the object of
  recent and fast-growing interest \cite{Luschen2017,VanNieuwenburg2018,Droenner2017,Xu2017}.

  {\it Systems considered.-}
  In this work, we focus on the transition from the weakly to the strongly correlated regime, and consider a single disorder strength.
  This specific choice is enough to display how the competition of disorder and interaction is captured within a DFT picture.
  We study quantum rings pierced by magnetic fields and electrically biased quantum-transport setups (Fig.~\ref{system}). Both situations show the aforementioned current crossover as function of the interaction strength. The rings are solved numerically exactly, while for quantum transport we use the Non-Equilibrium Green's Function (NEGF) formalism within many-body perturbation theory  \cite{Kadanoff1962,Keldysh1965,Haug2008,stefanucci2013,Balzer2013,Hopjan2014} to obtain steady-state currents and densities. The effective potentials and biases were found via a numerical reverse-engineering algorithm~\cite{Karlsson2016a} within non-equilibrium lattice DFT \cite{Verdozzi2008,Farzanehpour2012}.

  \section{Quantum rings}
  We study disordered Hubbard rings with $L=10$ sites, $N$ electrons, and spin-compensated, i.e. $N_\uparrow = N_\downarrow = N/2$. Currents are set by a magnetic field threading the rings, via the so-called Peierls substitution \cite{Peierls1933, Hofstadter1976}. The Hamiltonian is
  \begin{equation}
    \!\Hh = \!-T \! \! \sum _{\langle mn\rangle \sigma}\!\! e^{i\frac{\phi}{L}  x_{mn}} \chd_{m\sigma} \ch_{n\sigma}
    +\sum _{m\sigma}  (v_m
    \! +\! \frac{U}{2} \nh_{m,-\sigma}) \nh_{m\sigma},
    \label{PeierlsHamiltonian}
  \end{equation}
  where $\chd_{m\sigma}$ creates an electron with spin projection $\sigma=\pm 1$ at site $m$. $\nh_{m\sigma}=\chd_{m\sigma} \ch_{m\sigma}$ is the density operator, and $\langle ...\rangle$ denotes nearest-neighbor sites. $\phi$ is the Peierls phase and $x_{mn}=\pm 1$ depending on the direction of the hop from $m$ to $n$. $U$ is the onsite interaction. We consider onsite energies with box disorder of strength $W$ with $v_m \in \left [ -W/2, W/2 \right ]$.
  In passing, we note that Peierls phases can be realized experimentally in cold atoms by artificial gauge fields~\cite{Jimenez-Garcia2012}.
  We study currents in rings regimes via exact diagonalization, obtaining the many-body ground-state wavefunction $| \psi(\phi) \rangle$, and the corresponding density matrix $\rho_{mn} = \langle \psi (\phi)| \chd_{n\sigma} \ch_{m\sigma} | \psi (\phi)\rangle$.  This gives the density at site $m$ as $n_m = 2\rho_{mm}$ and the bond current as
  \mbox{$I_{m+1,m} = -4T \Im \left [ e^{i \phi /L} \rho_{m,m+1} \right ]$}.
  As we are in a steady-state scenario, all nearest-neighbor bond currents are equal, and the current $I\equiv I_{m+1,m}$ for any lattice site $m$.

  The corresponding effective KS Hamiltonian is~\cite{Akande2012,Akande2016}
  \begin{align}
    \hat{H}_{KS} =-T \sum _{\langle mn \rangle \sigma}  e^{i\frac{\phi^ {\KS}}{L}  x_{mn}}  \chd_{m\sigma} \ch_{n \sigma} +
    \sum _{m\sigma} v^{\KS}_m  \nh_{m\sigma}.
  \end{align}
  The $L+1$ effective parameters ($v^{\KS}_m, \phi^{\KS}$) are found by solving
  the KS equations $\left ( \Tb + \vb^\KS \right )\varphi_\nu=\epsilon_v \varphi_\nu$, where $(\Tb)_{mn}=-Te^{i\frac{\phi^ {\KS}}{L}  x_{mn}}$ for nearest neighbors and 0 otherwise, and $(\vb^\KS)_{mn} = \delta_{mn} v^\KS_m$.
  Imposing that the KS density $n_m = 2 \sum_{\nu = 1}^{N/2} \left | \varphi_\nu(m) \right |^2$ and KS bond current
  \mbox{
    $
    I  = -4T \sum_{\nu =1}^{N/2} \Im \left [ e^{i \phi^{\KS}/L} \varphi^*_\nu (m+1) \varphi_\nu (m)  \right ]
    $}
  equal those from the original interacting system determines ($v^{\KS}_m, \phi^{\KS}$). No physical  meaning should be \emph{a priori} given to the KS orbitals $\varphi_\nu$ or the KS eigenvalues $\epsilon_\nu$; they pertain to an auxiliary system giving the exact density and current but not necessarily other quantities.

  The KS potential, referred to as $v_{\eff}$ hereafter, is our proposed measure of disorder screening. It can be split into external (disorder) and Hartree-exchange-correlation parts: $v_{\eff} = v + v_{\Hxc}$ (similarly, $\phi_{\eff} \equiv \phi_{\KS} =  \phi + \phi_{\xc}$). Thus, in DFT, the screening of disorder by interactions (i.e. when $ |v_{\eff}| < |v| $) comes from $v_{\Hxc}$. This is an improvement over standard mean-field descriptions, in which the effective potential does not include correlations and the applied phase is unscreened.

  \begin{figure}[t]
    \centering
    \includegraphics[width=0.46\textwidth]{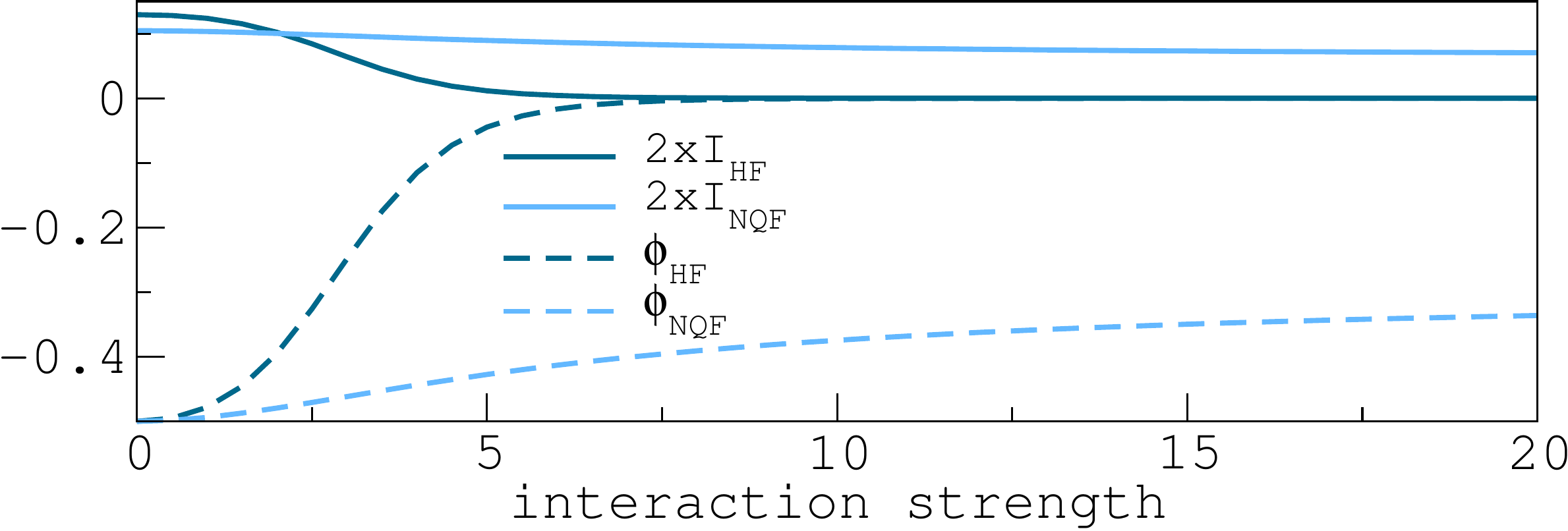}
    \caption{Current $I$ and KS phase $\phi_{\eff}$ in a 10-site homogeneous ring with density $n=3/5$ (NQF) and $n=1$ (HF).
      \label{homogeneousCase} }
  \end{figure}

Both $v_{\eff}$ and $\phi_{\eff}$ are obtained by mapping the exact many-body ring system into a DFT-KS one.
In lattice models, existence and uniqueness issues for such a DFT-based map can occur \cite{Baer2008,Li2008a,Verdozzi2008, Stefanucci2010,Kurth2011,Farzanehpour2012,Akande2012}.  Of relevance here, $\phi$ and $\phi + 2\pi k L$ ($k$ integer) give the same current (uniqueness issue); this  periodicity also implies that the magnitude of the current has an upper bound (existence issue). Further, a non-interacting (or described within KS-DFT)  homogeneous ring has energy degeneracy for even $N_\sigma$ (the degeneracy is  lifted by many-body interactions).

To circumvent these occurrences, we choose $N_\sigma$ odd to avoid degeneracies. Furthermore, we fix $-\pi/L < \phi_\eff \leq \pi/L$ in the reverse engineering scheme. However, even with this restriction, two different phases can yield the same current. Practically, we consistently choose the region for $\phi_{\eff}$ that smoothly connects to $\phi$ for small $U$.
Finally, in practice the "maximal current'' existence issue is largely mitigated since the target current comes from a physical many-body system.

  In the numerical reverse-engineering implementation of the DFT map, $\phi_{\eff}$ and $v_{\eff}$ are recursively updated until the interacting MB system and the KS system have the same current and density.
  Using exact diagonalization, we obtain the exact many-body density $n_{\text{MB}}$ and current $I_{\text{MB}}$. These quantities are then used as input to obtain $\phi_{\eff}$ and $v_{\eff}$ according to the protocol~\cite{Karlsson2016a}
  \begin{align}
    &v_{\eff}^{(k+1)} = v_\eff^{(k)} + \alpha_1
    \left (n^{(k)}_{\text{KS}} - n_{\text{MB}}\right ) \quad \text{   for all sites} \label{iterativeDensity} \\
    &\phi_{\eff}^{(k+1)} =
    \phi_{\eff}^{(k)} - \alpha_2 \left (I^{(k)}_{\text{KS}} -
I_{\text{MB}}\right ), \label{iterativeCurrent}
  \end{align}
  where $(k)$ denotes the $k$th iteration, and $\alpha_1,\alpha_2 < 1$ are convergence parameters.

\subsection{Quantum rings: results}
  We consider two electron concentrations: half-filling (HF, $N_\uparrow = 5$), and near quarter-filling (NQF, $N_\uparrow= 3$). Furthermore, we take $T=1$, i.e. the energy unit.

  For reference, we start our discussion with homogeneous rings, (i.e. $v_i = 0$, which gives a constant density $n_i=N/L$ and a constant $v_{\eff}$). In Fig.~\ref{homogeneousCase}, we show HF and NQF currents and the corresponding $\phi_{\eff}$:s as function of the interaction $U$, for fixed external phase $\phi = -0.5$. Both HF and NQF currents $I$ decrease monotonically with $U$, but tend to zero and nonzero values, respectively. This is consistent with Mott insulator behavior at HF and metallic behavior otherwise for the infinite $(L \to \infty)$ one-dimensional Hubbard model~\cite{Lieb1968}.
  The homogeneity singles out the importance of the effective phase. The KS orbitals are plane waves for any value of $U$, and as such the current is determined solely by $\phi_{\eff}$. This shows the importance of the effective phase in our Hamiltonian picture, and highlights that standard mean-field descriptions, which yields $\phi_\eff = \phi$, cannot capture the correct physics.

  We now address the effect of disorder in rings. We use $M=150$ box-disorder configurations. For a given configuration, the spread $\Delta X$ of a quantity $X$ over the $L=10$ sites is measured by
  \begin{equation}
    (\Delta X)^2 = \frac{1}{L}\! \sum _{m=1}^L \! \left ( \bar{X} - X_m \right )^2\!, \text{ with } \bar{X} = \frac{1}{L}\sum _{m=1}^L X_m. \label{averageDefinition}
  \end{equation}
  Results are presented for i) histograms collecting data from each disorder configuration and ii) arithmetic averages over all $M$ configurations. We examine the dependence on the interactions $U$ of the current $I$, $\phi_{\eff}$, $\Delta n$,and $\Delta v_{\eff}$. The latter is a measure of disorder screening (in the homogeneous case, $\Delta v_{\eff} = 0$ for all $U$).

\begin{figure}[t]
    \centering
    \includegraphics[width=0.48\textwidth]{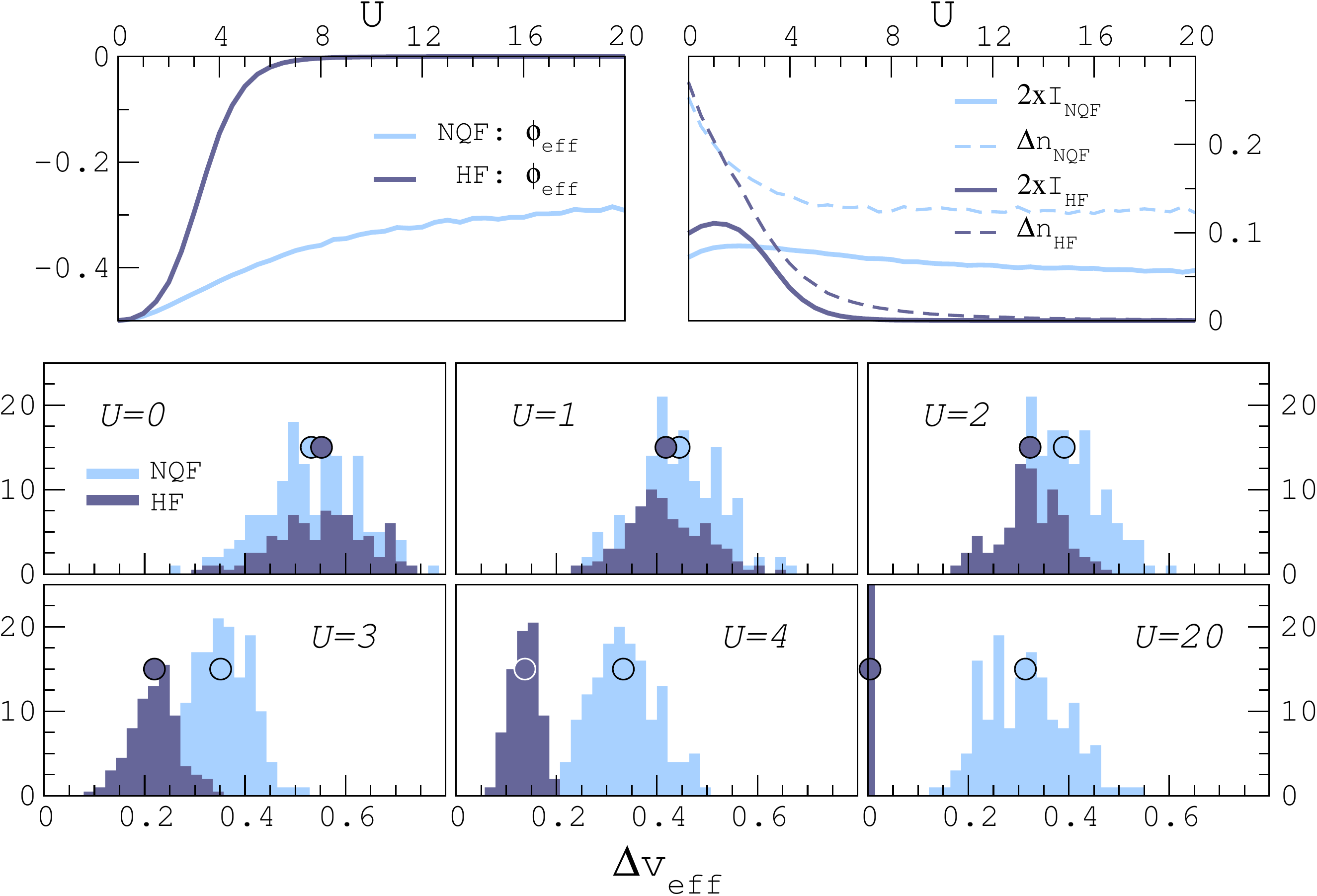}
    \caption{Disorder vs interactions in 10-site rings near quarter-filling (NQF, $N=6$) and at half-filling (HF, $N=10$) for $W=2$, $\phi=-0.5$. For $\Delta v_{\eff}$, histograms and disorder averages are shown.
      For $\phi_{\eff}$, $I$, $\Delta n$ disorder averages are reported.
      \label{10SitesNQF} }
  \end{figure}

  With disorder ($W=2$), for both NQF and HF the current $I$ is hindered by disorder at low $U$ and by interactions at large $U$, with a maximum in between (Fig.~\ref{10SitesNQF}).
  As for $W=0$ (Fig.~\ref{homogeneousCase}), for HF $I$ vanishes at very large $U$. The non-monotonic behavior of $I$ results from competing disorder and interactions \cite{Farhoodfar2011,Vojta1998,Karlsson2014b}. Conversely, the density spread $\Delta n$ decreases monotonically as a function of $U$ at both NQF and HF, i.e. interactions favor a more homogeneous density. For NQF, $\Delta n$ seems to tend to a finite value for large $U$, while for the HF case, $\Delta n \to 0$, i.e. a fully homogeneous density.
  In the KS system, $\Delta v_{\eff}$ also decreases monotonically as function of $U$, tending to a finite value for NQF and to zero for HF.
  This means that  the \emph{exact} $v_{\eff}$ for a strongly correlated system is smoother than for a weakly correlated system, and similarly for the density. Thus, we cannot simply look at the spread of the density to predict the current through the system: Including the effective phase is crucial.

  The competition of disorder and interactions thus translates into a competition of a decreasing effective potential spread (favoring the current) and a decreasing effective phase (reducing the current). Mean-field \cite{Farhoodfar2011} or DFT-LDA treatments \cite{Vettchinkina2013} fail to explain the current drop since they only take the effective potential into account: With an effective potential and no effective phase, the current can only increase with interactions. This ends our discussion on exact treatments of quantum rings.

  \section{Open systems}
  We study short clusters connected to semi-infinite leads, with Hamiltonian
  \begin{equation}
  \Hh = \Hh _{C} + \Hh _{l} + \Hh _{Cl},\label{Ham}
  \end{equation}
  where $C$, $l$, and $Cl$ label the cluster, leads, and cluster-leads coupling  parts, respectively. With the same notation as for rings,
  \begin{align}
    \Hh _{C} \! = \! -T \! \! \! \! \! \! \! \sum_{\langle mn \rangle  \in C, \sigma } \! \! \! \! \! \! \! \chd_{m\sigma} \ch_{n\sigma}  + \sum _{m \sigma} \! v_m \nh_{m\sigma} + U\sum _m \! \nh_{m\uparrow} \nh_{m\downarrow}.
    \label{central}
  \end{align}
  As in the case of the quantum rings, we consider box disorder, $v_m \in [-W/2, W/2]$. Depending on the cluster dimensionality, the leads are either 1D (chain) or 2D (strip) semi-infinite tight-binding structures. The latter case is shown in
  Fig.~\ref{system}.
  The lead Hamiltonian is  $\Hh_{l}=\sum_\alpha \Hh_{\alpha}$, and $\alpha= r (l)$ refers to the right (left) contact:
  \begin{equation}
   \Hh_{\alpha } = -T \! \! \! \! \! \sum _{\langle mn \rangle \in \alpha, \sigma} \! \! \! \!  \chd_{m\sigma} \ch_{n\sigma} + b_\alpha (t) \hat{N}_\alpha.
  \end{equation}
  Here, $b_\alpha(t)$ is the (site-independent) bias in lead $\alpha$, and  $\hat{N}_\alpha = \sum _{m \in \alpha,\sigma} \nh _{m\sigma}$ the number operator in lead $\alpha$.
  Finally, the lead-cluster coupling $\Hh_{Cl}$ connects the edges of the central region to the leads  (Fig.~\ref{system}) with tunneling parameter $-T$. In the following, we put $T=1$, which defines the energy unit. We focus on the steady-state scenario with $b_r (t) = 0$, and $b_l \equiv b_l(t\to\infty)=1$, beyond the linear regime. Our 1D and 2D clusters have $L=10$ sites, but are large enough to illustrate the relevant physics and the scope of a DFT perspective. Also, we put $n_\uparrow = n_\downarrow = n$ (non-magnetic case) and the temperature to zero.

  \subsection{Steady-state Green's functions}
  Both our many-body (MB) and KS treatments of open systems are based on NEGF in its steady-state formalism. Thus we keep our presentation general, and later specialize to MB or KS. To describe the steady-state regime, we use retarded $\Gb^R(\omega)$ and lesser $\Gb^<(\omega)$ Green's functions:
  \begin{align}
    &\Gb^R(\omega) = \left [~\omega {\bm 1} - \Tb - {\bm  v} - \Sigmab^R (\omega)~ \right ]^{-1},\label{eq:retG}\\
    &\Gb^<(\omega) = \Gb^R(\omega) \Sigmab^< (\omega) \Gb^A(\omega). \label{eq:lesserG}
  \end{align}
  Here, boldface quantities denote $L \times L$ matrices in site indices
  of the cluster region. $\Gb^A = (\Gb^R)^\dagger$ is the advanced Green's function,  $(\Tb)_{mn} = T_{mn}$ is the kinetic term of \Eq{central}, and ${\bm  v}$ is not specified yet. The self-energy $\Sigmab$ contains many-body (MB) and embedding (emb) parts: $\Sigmab^{R/<}=\Sigmab^{R/<} _{\MB} + \Sigmab^{R/<} _{\text{emb}}$.
  All correlation effects are contained in the many-body self-energy $\Sigmab_{\MB}^{R/<}$, whilst the embedding term
  accounts in an exact way for the left ($l$) and right ($r$) leads~\cite{Myohanen2008}:
  $\Sigmab_{\text{emb}}^{R/<}= \sum_{\alpha=l,r} \Sigmab_{\alpha}^{R/<}$. More explicitly, $ \Sigmab_{\alpha}^{<}(\omega)=i f_\alpha(\omega) \Gammab ^\alpha (\omega)$, where $\Gammab^\alpha = -2\Im \left [\Sigmab^R_{\alpha}\right ]$ and
  $ f_\alpha(\omega)=\theta(-\omega + \mu + b_\alpha)$. Thus, information of the
  actual structure of the leads, the bias $b_\alpha$ and the chemical potential $\mu$ enters via $f_\alpha$ and $\Sigmab_{\text{emb}}^{R}$. Explicit expressions of $\Sigmab^R_{\text{emb}}$ exist for 1D and 2D\cite{Myohanen2012Thesis} semi-infinite leads, since they are determined by the uncontacted-lead case.

  For our system, the steady-state particle density and current are spin-independent. In each spin channel \cite{Meir1992}, with $I_l$ the left lead current, and the spin-labels omitted,
  \begin{align}
    &n_k =  \int _{-\infty} ^\infty \frac{d\omega}{2\pi i} G_{kk}^<(\omega), \label{eq:density}\\
    &I_l =\!\! \int _{-\infty} ^\infty \! \frac{d\omega}{2\pi i}
    \Tr \left [\Gammab ^l (\omega) \left ( \Gb^< (\omega) - 2\pi i f_l(\omega) {\bm A} (\omega)   \right )     \right ], \label{eq:meir}
  \end{align}
  where the spectral function is $ 2\pi {\bm A} = i(\Gb^R-\Gb^A)$. Both the interacting MB system as well as the KS system are described by Eqs.~(\ref{eq:retG}-\ref{eq:meir}). We now specialize the discussion to the separate cases.

  \begin{figure}[t]
    \centering
    \includegraphics[width=0.48\textwidth]{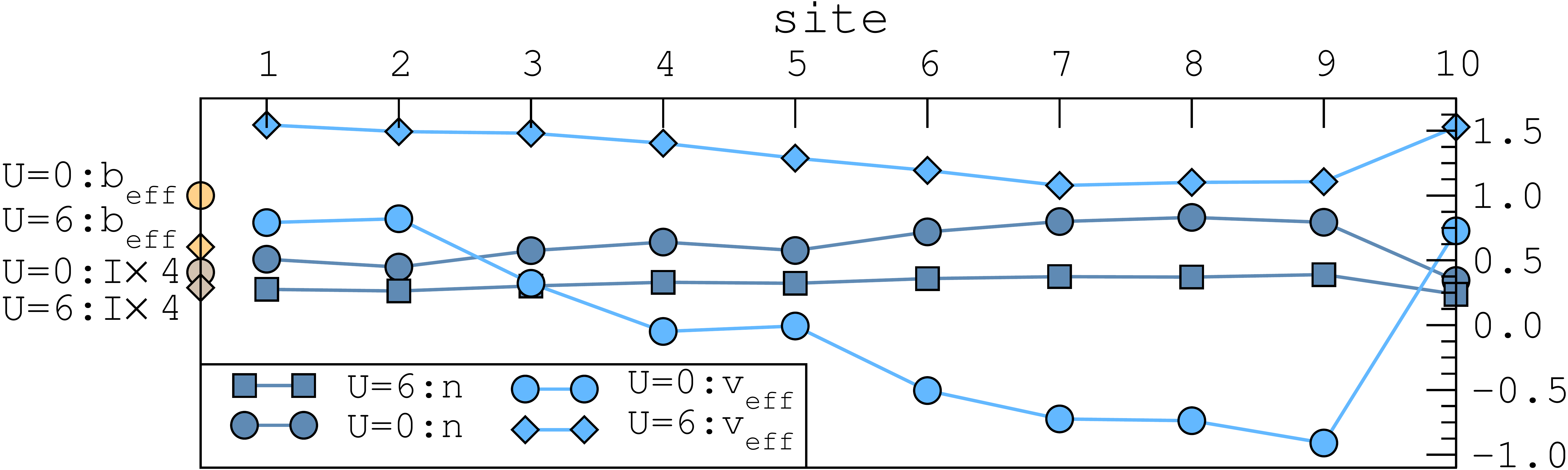}
    \caption{Density $n$, effective potential $v_{\eff}$, effective bias $b_{\eff}$ and current $I$ (multiplied by 4 for convenience) for a specific one-dimensional disordered wire with $W=2$ and bias $b=1$  for $U=0$ and $6$.
      \label{oneConfiguration} }
  \end{figure}

  \subsubsection{The interacting MB system}
  Here $(\bm  v)_{ij}=\delta_{ij}v_i$ are the disordered onsite energies and $b_\alpha$ is the applied bias.
  While the NEGF formalism provides a formally exact description for open systems, in practice the MB self-energies $\Sigmab_{\MB}^{R/<}$ need to  be approximated. We consider the self-consistent $\Sigmab_{\MB}^{R/<} = \Sigmab_{\MB}^{R/<} [\Gb^R,\Gb^<]$ 2nd Born approximation, \cite{Myohanen2008,Friesen2009,Karlsson2014b,Thygesen2008}, keeping all diagrams up to second order. While the numerical details depend on the chosen approximation, our conclusions do not, as discussed in more detail below.
  We solve the equations self-consistently, with the convergence rate improved with the Pulay scheme \cite{Pulay1980,Thygesen2008}. Fully self-consistent NEGF calculations guarantee the satisfaction of general conservation laws \cite{Baym1961, Baym1962}, and in particular the continuity equation \cite{Karlsson2016c}. In the context of steady-state transport, the continuity equation leads to the condition that $I_l = -I_r \equiv I$.

  \subsubsection{The independent-particle KS system}
  Being an independent-particle system, $\Sigmab_{\MB}^{R/<}=0$. Thus, the KS system is described exactly by steady-state NEGF. Further, $(\vb)_{ij} \equiv \delta_{ij} v_{i,\eff}$ and $b_{\alpha} \equiv b_{\eff,\alpha}$ are found iteratively to make the KS and MB density and current the same \cite{Karlsson2016a}. The same iteration protocol as for the quantum rings was used, \Eq{iterativeDensity} and \Eq{iterativeCurrent}, replacing $\phi_\eff$ with $b_\eff$.
  The embedding self-energies in the KS and MB systems differ only by the bias (effective in KS, applied in MB). We restrict $b_{\eff,r} = 0$ and define $b_{\eff} = b_{\eff,l}$.

  The KS independent-particle scheme permits to
  write the Meir-Wingreen formula, \Eq{eq:meir}, in a Landauer-B\"uttiker form
  $I = \int _{\mu}^{\mu + b_{\eff}} \frac{d\omega}{2\pi} \mathcal{T}_{KS} (\omega)$, with $\mathcal{T}_{KS} = \Tr \left [  \Gammab^l \Gb^R\Gammab^r \Gb^A \right ]$ the KS transmission function. Although recast in a Landauer-B\"uttiker form used for independent-particle systems, we stress that the KS current still equals the true current of the original interacting system.

  \subsection{Open systems: results}
  In the following, we put $\mu_\alpha = 0$ (half-filled leads). To address the behavior of $v_{\eff}$  in quantum transport setups, we find it useful to start with one disorder configuration and two interaction values for a biased 10-site one-dimensional chain (Fig.~\ref{oneConfiguration}). At $U=0$, the density $n_k$ is non-uniform, since $v_{\eff} = v$ and $b_{\eff} = b$. For $U=6$, both $n_k$ and $v_{\eff}$ (now incorporating correlations) become smoother: interactions thus provide a smoother energy landscape also for open systems. However, $I_{U=6}< I_{U=0}$, even if the effective energy  landscape is smoother. This is due to $b_{\eff}$, that at $U=6$ is much smaller than $b$ \cite{Stefanucci2004a,Stefanucci2015,Karlsson2016a}.

\begin{figure}[t]
    \centering
    \includegraphics[width=0.47\textwidth]{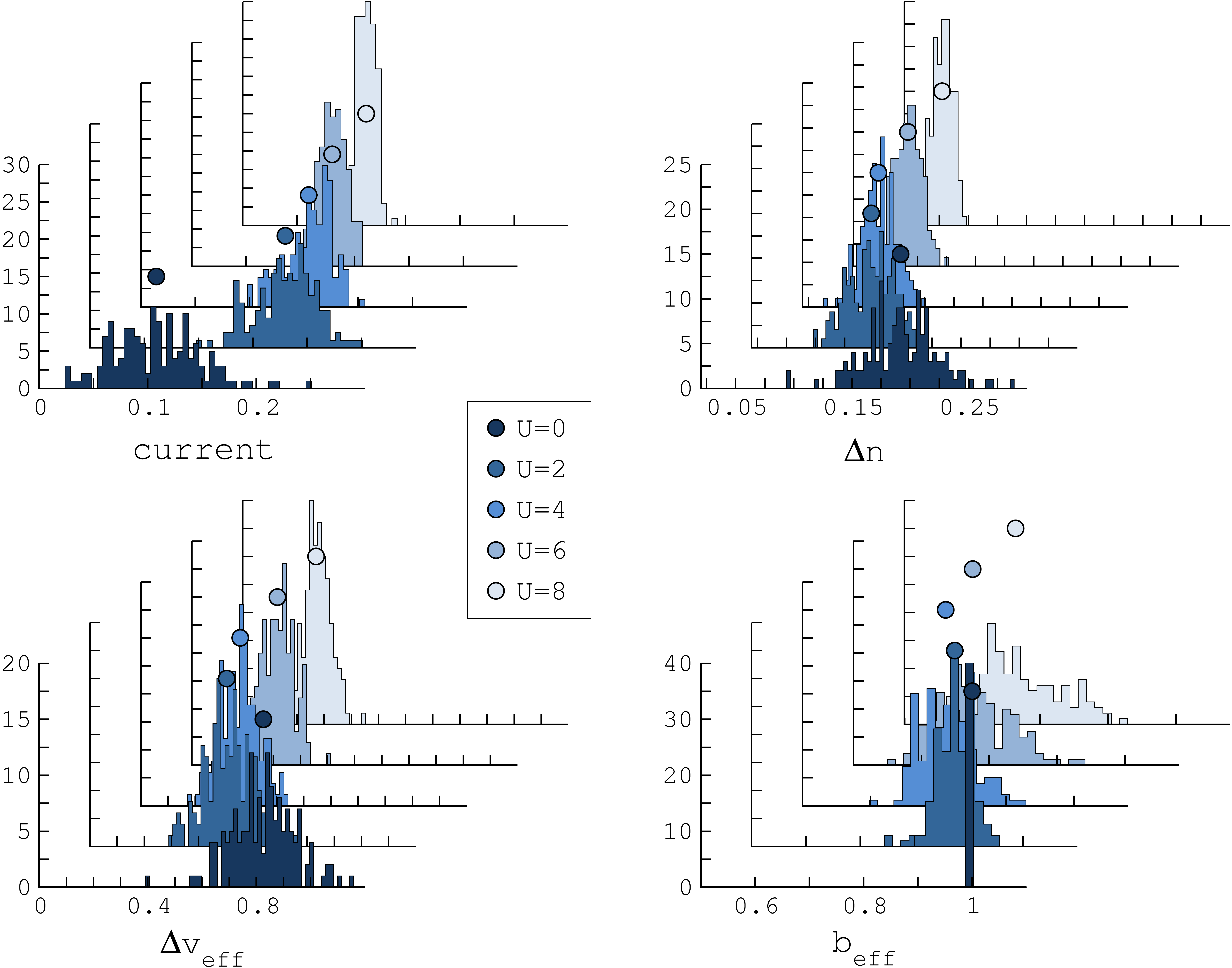}
    \caption{Histograms of $\Delta n$, $\Delta v_{\eff}$, $I$ and $b_{\eff}$ and their arithmetical averages (dots) for the 2D quantum transport system of Fig.~\ref{system} with disorder strength $W=3$ and bias $b=1$.
      The corresponding statistical errors $\sigma_x = \frac{\bar{x}}{\sqrt{M}}$ are comparable or smaller than the dot sizes, and thus not shown. Results for the 1D open system exhibit similar trends.}
    \label{histogram2D}
  \end{figure}

  To corroborate this analysis, we consider the 2D open system of Fig.~\ref{system}. Results from 150 disorder configurations for  $\Delta n$, $\Delta v_{\eff}$ (defined as for rings, \Eq{averageDefinition}), $I$, and $b_{\eff}$ are shown in Fig.~\ref{histogram2D}
  \footnote{We do not include results from a small fraction of calculations $(\sim 1\%)$, which, especially for large $W$, did not converge.}.
  The current through the system is a non-monotonic function of $U$, while
  $\Delta n$, $\Delta v_{\eff}$ decrease monotonically. At low $U$, $b_{\eff}$ almost equals $b$, and $I$ increases since $\Delta v_{\eff}$ decreases. At larger $U$, however, the drop in $b_{\eff}$ grows, and $I$ is smaller. This is why $I$ shows a crossover.
  Thus, the competition between disorder and interactions in open systems transfers to a competition between the smoothness of the energy landscape favoring current flow and screening of the effective bias hindering such flow. We have performed the same analysis for one-dimensional linear chains, with the same qualitative results (not shown).

  While the 2nd Born approximation is quite accurate at low interaction strengths \cite{Friesen2009,Uimonen2011, Hermanns2014a}, one can of course question the quantitative agreement for higher interaction strengths. We find no reason to question the qualitative results of the approximation, however, since the behavior is similar for the quantum rings, which were treated exactly, and also other calculations suggest similar conclusions\cite{Denteneer1999,Vojta1998,Abrahams2001}. In order to further confirm the aforementioned qualitative behavior, we also performed calculations for selected disorder configurations (not shown) using the T-matrix approximation~\cite{Galitskii1958,Friesen2009,PuigvonFriesen2011,Schlunzen2016}, which takes higher-order processes into account in the self-energy. We found the same qualitative behavior as for the 2nd Born approximation, also for the higher interaction strengths considered.

  \section{Conclusions}
  We introduced an exact independent-particle characterization of coexisting disorder and interaction effects, based on density-functional theory (DFT). Its scope as a diagnostic for disorder screening was shown for open-sample geometries and small quantum rings.
  The many-body treatment of the quantum rings was exact, which allowed us to unambiguously characterize disorder screening. For open systems, where no exact solutions are available, we used non-equilibrium Green's functions (NEGF), with biased reservoirs treated exactly and electronic correlations treated via the 2nd Born approximation. We stress that the use of an approximation was simply an expedient way to provide an input to our reverse engineering algorithm; more sophisticated methods can of course be used for the same purpose.

  Our DFT-based analysis consistently shows that interactions smoothen the energy landscape in disordered systems out of equilibrium, for both closed quantum rings and open one- and two-dimensional quantum transport systems. In line with earlier qualitative pictures from the literature, it is tempting to think that the spread in the effective potential or the density can be taken as a measure of the conductance of a system. This is not the case, as this picture is not accurate enough to explain the non-monotonic behavior for the current when changing the interaction strength. To make the picture complete, the effective bias (phase) has to be taken into account.
  The fact that the quantum rings, the one-dimensional and the two-dimensional quantum transport systems yield the same behavior reinforces our conclusions that the independent-particle picture is general and can be applied to a wide range of systems.

  Based on this interpretation, we can provide a simple explanation why mean-field theories can predict a too high current in disordered systems. These methods neglects the correlation screening of the disordered potential, and fully neglects the screening of the applied bias. To improve the picture, correlation effects need to be added.

  To conclude, within our Hamiltonian independent-particle picture, strong correlation effects are behind the appearance of the effective  potential and bias, and this is the essence of disorder screening.
  As possible extensions of our approach, we mention applications to real materials and the generalization to finite temperatures~\cite{Mermin1965} to describe, for example, the many-body localized regime.
  These are deferred to future work.

  \begin{acknowledgments}
    We acknowledge Fabian Heidrich-Meisner for useful discussions. D.K. would like to thank the Academy of Finland for support under project no. 267839.
  \end{acknowledgments}

%
\end{document}